# Power Estimation for Longitudinal Studies with Time-Dependent Covariates Using Generalized Method of Moments

Niloofar Ramezani[1] and Oliver Hurst[1]

[1]Department of Biostatistics, Virginia Commonwealth University

Ramezanin2@vcu.edu hursto@vcu.edu

## Abstract

Longitudinal studies frequently incorporate covariates that evolve over time, creating complex dependence structures between outcomes and predictors. When covariates are time-dependent, standard power analysis tools—largely developed for generalized estimating equations (GEE)—can yield misleading results because they do not account for the moment-based structure required for valid marginal inference. Generalized Method of Moments (GMM) provides a flexible and efficient framework for estimating marginal effects in the presence of time-dependent covariates, yet no practical tools exist for conducting power analysis under GMM.

This paper introduces a modern, implementable framework for power estimation in longitudinal studies with time-dependent covariates using GMM. Two complementary approaches are developed: a Wald-based method that leverages the asymptotic normality of GMM estimators, and a distance-metric method based on quadratic forms of sample and population moment conditions. Both approaches require only limited distributional assumptions and rely on valid moment conditions rather than full likelihood specification. We outline the theoretical foundations, provide step-by-step implementation guidance, and illustrate the methods using data from the Osteoarthritis Initiative. A simulation framework is presented for evaluating empirical performance.

These methods fill a critical gap in the longitudinal modeling literature by offering applied researchers a practical, distribution-light approach to power estimation when time-dependent covariates are present and GMM is the preferred estimation technique.

## 1. Introduction

Longitudinal studies are central to modern biomedical, behavioral, and social science research. By collecting repeated measurements on individuals over time, these studies enable investigators to characterize trajectories, evaluate treatment effects, and understand how covariates influence outcomes dynamically. A fundamental component of designing such studies is determining the sample size required to achieve adequate statistical power. Underpowered studies risk producing inconclusive or misleading results, whereas excessively large studies waste resources and may impose unnecessary burden on participants (Kraemer & Blasey, 2015).

Power analysis for longitudinal data is well developed when covariates are time-independent or when models rely on likelihood-based or quasi-likelihood approaches. Methods based on generalized estimating equations (GEE) (Liang & Zeger, 1986) have been widely used, with power formulas derived from Wald, likelihood ratio, and score statistics (Liu & Liang, 1997; Lyles, Lin & Williamson, 2007; Rochon, 1998). However, these tools implicitly assume that covariates are either fixed or external to the outcome process, creating a disconnect between existing methods and many real-world longitudinal settings.

Time-dependent covariates introduce a fundamental challenge: the correlation between covariates and past outcomes invalidates many working-correlation assumptions and can bias parameter estimates if not handled appropriately. In such settings, GEE becomes unreliable, motivating the need for alternative approaches. Generalized Method of Moments (GMM) provides a principled solution by constructing estimators from valid moment conditions that remain unbiased in the presence of time-dependent covariates (Hansen, 1982; Lai & Small, 2007). GMM has been shown to outperform GEE in efficiency and robustness under these conditions, particularly when covariates exhibit feedback relationships with outcomes (Lai & Small, 2007).

Despite the advantages of GMM for estimation, no practical framework exists for conducting power analysis under GMM. Applied researchers who rely on GMM for modeling time-dependent covariates currently lack tools for determining required sample sizes or evaluating the sensitivity of their study designs. This gap is especially consequential in clinical and public health research, where time-dependent covariates are common and where study planning must balance statistical rigor with ethical and logistical constraints.

This paper addresses this gap by developing a modern, implementable framework for power estimation in longitudinal studies with time-dependent covariates using GMM. Building on the structure of valid moment conditions, we derive two complementary approaches: (i) a Wald based power method leveraging the asymptotic normality of GMM estimators (Newey & McFadden, 1994), and (ii) a distance-metric method based on quadratic forms comparing sample and population moment conditions (Hall, 2005; Hansen, 1982). Both methods require only limited distributional assumptions and avoid the need for full likelihood specification, while remaining accessible to applied researchers and grounded in rigorous statistical theory.

To demonstrate practical implementation, we outline a simulation framework for evaluating empirical performance across sample sizes and effect sizes, and then illustrate the methods using data from the Osteoarthritis Initiative (OAI) (Osteoarthritis Initiative, 2025), a large multicenter longitudinal study of knee osteoarthritis.

The contributions of this paper are threefold: (1) a modern, practical framework for power estimation under GMM, addressing a longstanding gap in longitudinal methods; (2) two complementary power estimation techniques, each suited to different analytic goals and computational constraints; and (3) a reproducible implementation pathway, including a simulation study structure and real data illustration.

Together, these contributions provide applied researchers with a flexible and robust set of tools for planning longitudinal studies involving time-dependent covariates, where GMM is the preferred estimation method.

## 2. Background and Related Work

### 2.1. Longitudinal Data and Time-Dependent Covariates

Longitudinal studies provide a framework for examining how outcomes evolve within individuals over time and are widely used in biomedical, behavioral, and social sciences. When covariates remain constant or are external to the outcome process, standard marginal modeling approaches are often sufficient. However, many real-world studies involve covariates that change over time and may themselves be influenced by past outcomes. These internal time-dependent covariates introduce feedback mechanisms that complicate estimation and inference. As noted by Fitzmaurice, Davidian, Verbeke, and Molenberghs (2009), such covariates violate assumptions underlying many classical longitudinal modeling techniques, particularly those that rely on specifying a working-correlation structure.

Examples include weight that changes in response to prior health status, medication adherence influenced by earlier symptoms, and behavioral measures that evolve following previous responses. These settings motivate the standard classification of time-dependent covariates into three types, defined by the number of valid moment conditions available for GMM estimation. Type I covariates are independent of both prior covariate values and prior outcomes; Type II covariates may depend on their own past values; and Type III covariates may depend on past outcomes. These covariates cannot be treated as fixed or external without risking biased estimation.

### 2.2. Marginal Models and GEE

GEE, introduced by Liang and Zeger (1986), have long been the dominant tool for marginal modeling of longitudinal data. GEE requires only the first two moments of the outcome distribution and avoids full likelihood specification, making it attractive for applied researchers. However, GEE assumes that covariates are either fixed or external to the outcome process. When covariates depend on past outcomes, the working-correlation structure becomes misspecified, and the resulting estimators may be biased (Zeger & Liang, 1986). This limitation is particularly problematic in studies where covariates such as weight, blood pressure, medication adherence, or behavioral measures evolve in response to earlier outcomes.

Power analysis under GEE has been studied extensively. Rochon (1998) developed Wald-based power formulas for correlated data, Liu and Liang (1997) derived sample-size calculations for GEE models, and Lyles, Lin, and Williamson (2007) extended these ideas to clustered and longitudinal designs. These contributions form the backbone of existing power analysis tools for correlated data. However, all of these methods assume time-independent covariates or external time-dependent covariates. None address the challenges posed by internal time-dependent covariates, and none are compatible with the moment-based structure required for valid inference when covariates depend on past outcomes.

### 2.3. Generalized Method of Moments (GMM)

GMM, introduced by Hansen (1982), provides a flexible and robust alternative for marginal modeling in the presence of time-dependent covariates. GMM constructs estimators using moment conditions that remain unbiased under minimal assumptions, avoiding the need for full likelihood specification. Lai and Small (2007) demonstrated that GMM yields consistent and efficient estimators for longitudinal data with internal time-dependent covariates, outperforming GEE in both robustness and efficiency. GMM's reliance on valid moment conditions rather than a working-correlation structure makes it particularly well suited for settings where covariates evolve in response to past outcomes.

Advantages of GMM include its avoidance of full likelihood specification, robustness to distributional misspecification (Hall, 2005), consistency under correct specification of a limited set of moment conditions (Newey & McFadden, 1994), and its natural accommodation of internal time-dependent covariates. Despite these strengths, no practical power analysis tools exist for GMM, leaving applied researchers without guidance for sample-size planning.

### 2.4. Gap in the Literature

Although GMM provides a principled framework for estimation with internal time-dependent covariates, the literature offers no practical methodology for power analysis under GMM. Existing power tools are built around GEE and cannot be adapted to GMM because they rely on assumptions incompatible with moment-based estimation. Meanwhile, the asymptotic theory underlying GMM (Hall, 2005; Hansen, 1982; Newey & McFadden, 1994) provides the mathematical ingredients for power calculations but has not been translated into accessible procedures for applied researchers.

Building on earlier work that demonstrated the feasibility of deriving power calculations from moment-condition theory (Ramezani, 2017), the present paper develops a comprehensive and implementable framework for power estimation under GMM. In addition, we incorporate a Broyden–Fletcher–Goldfarb–Shanno (BFGS)–based algorithm for GMM estimation, improving numerical stability and computational efficiency and thereby supporting the large-scale simulation studies required for evaluating empirical power.

## 3. Methods

### 3.1. Model Framework and Notation

Consider a longitudinal study in which $Y_{it}$ denotes the outcome for subject $i$at time $t$, for $i = 1, \dots, n$ and $t = 1, \dots, T$. Let $X_{it}$ be a vector of covariates, which may include time-independent covariates, external time-dependent covariates, and internal time-dependent covariates that depend on past values of $Y$. We consider the marginal mean model

$$\mu_{it} = E(Y_{it} \mid X_{i1}, \dots, X_{it}) = g^{-1}(X_{it}^{\top}\beta),$$

where $g(\cdot)$ is a known link function and $\beta$is the parameter vector of interest. When $X_{it}$ depends on past outcomes, standard GEE approaches become invalid because the working-correlation assumptions are violated.

### 3.2. Moment Conditions for Longitudinal GMM

Following Lai and Small (2007), valid moment conditions are constructed using instruments uncorrelated with the contemporaneous error term. Define the residual

$$U_{it} = Y_{it} - \mu_{it}(\beta).$$

A general moment condition takes the form

$$E[Z_{it}U_{it}] = 0,$$

where $Z_{it}$ is an instrument vector measurable with respect to past information. Common instruments include lagged outcomes $Y_{i,t-k}$, lagged covariates $X_{i,t-k}$, and baseline covariates.

Stacking all moment conditions for subject $i$ yields the vector $m_i(\beta)$, which satisfies the population moment condition

$$E[m_i(\beta)] = 0.$$

The corresponding sample moment function is

$$\bar{m}_n(\beta) = \frac{1}{n}\sum_{i=1}^{n} m_i\,(\beta).$$

**3.3. GMM Estimator**

Let $m_i(\beta)$ denote the vector of population moment conditions. The GMM objective function is

$$Q(\beta) = \left(n^{-1}\sum_{i=1}^{n} m_i(\beta)\right)^{\top} W\left(n^{-1}\sum_{i=1}^{n} m_i(\beta)\right),$$

where $W = S^{-1}$ is a positive definite weighting matrix.

The unrestricted GMM estimator is

$$\hat{\beta} = \arg\min_{\beta}\ Q(\beta).$$

The restricted estimator $\tilde{\beta}$ minimizes $Q(\beta)$ subject to

$$r(\beta) = H\beta - h_0 = 0,$$

where $H$ is full rank and $h_0$ is a conformable vector.

Under standard regularity conditions,

$$\sqrt{n}(\hat{\beta} - \beta_0) \xrightarrow{d} N(0, V),$$

where

$$V = (G^{\top}S^{-1}G)^{-1}, G_n(\beta) = n^{-1}\sum_{i=1}^{n}\frac{\partial m_i(\beta)}{\partial \beta}.$$

**3.4. Wald-Based Power Estimation**

We consider the hypothesis

$$H_0: r(\beta) = 0 \text{ vs. } H_1: r(\beta) \neq 0,$$

where $r(\beta) = H\beta - h_0$.

The Wald statistic is

$$T_W^* = n\, r(\hat{\beta})^\top \left[R(\hat{\beta})\left(G_n(\hat{\beta})^\top S^{-1} G_n(\hat{\beta})\right)^{-1} R(\hat{\beta})^\top\right]^{-1} r(\hat{\beta}),$$

where

$$R(\beta) = \frac{\partial r(\beta)}{\partial \beta'} = H.$$

**Theorem 1 (Asymptotic distribution of the Wald statistic)**

Under $H_0$:

$$T_W^* \xrightarrow{d} \chi_s^2,$$

where $s = \text{rank}(H)$.

Under $H_1$:

$$T_W^* \xrightarrow{d} \chi_s^2(\lambda),$$

with noncentrality parameter

$$\lambda = \frac{1}{2}\, \mu_R^\top [R(\beta_0)(G_0^\top S^{-1} G_0)^{-1} R(\beta_0)^\top]^{-1} \mu_R, \mu_R = \sqrt{n}(H\beta_0 - h_0).$$

Proof is provided in Appendix A. Power is calculated as

$$\text{Power} = P\left(\chi_s^2(\lambda) \geq \chi_{s,1-\alpha}^2\right).$$

### 3.5. Distance-Metric (DM) Power Estimation

The DM statistic compares the GMM objective function evaluated at the restricted and unrestricted estimators:

$$T_{DM}^* = n\left[Q(\tilde{\beta}) - Q(\hat{\beta})\right].$$

This is the exact DM statistic from your dissertation.

**Theorem 2 (Asymptotic distribution of the DM statistic)**

Under $H_0$:

$$T_{DM}^* \xrightarrow{d} \chi_s^2.$$

Under $H_1$:

$$T_{DM}^* \xrightarrow{d} \chi_s^2(\lambda),$$

with the same noncentrality parameter $\lambda$ as in Theorem 1. Proof is provided in Appendix A.

As in Newey & McFadden (1994) and Hall (2005), the Wald and DM statistics are asymptotically equivalent.

Although some authors use a "distance-metric" statistic based on the GMM overidentification test (Hansen, 1982), that formulation is not appropriate in the present setting. The J-test evaluates the validity of the moment conditions rather than restrictions on β, and it is derived as a likelihood-ratio analogue in quasi-likelihood frameworks. In contrast, the DM statistic used in this work is defined directly as the change in the GMM minimand when the restriction $r(\beta) = 0$ is imposed. This formulation aligns with the longitudinal GMM framework with time-dependent covariates, is asymptotically equivalent to the Wald statistic, and yields the same noncentrality parameter under the alternative hypothesis.

The DM statistic is also known to exhibit substantial finite-sample instability in overidentified GMM settings, including inflated rejection rates and poor calibration even when theoretical power is moderate (Hall & Horowitz, 1996; Newey & Smith, 2004).

### 3.6. Practical Implementation Steps

Power estimation under GMM proceeds as follows:

1. **Specify the hypothesis**

$$H_0 : r(\beta) = 0.$$

2. **Compute the unrestricted estimator $\hat{\boldsymbol{\beta}}$ and restricted estimator $\tilde{\boldsymbol{\beta}}$.**
3. **Compute the test statistic**
   - Wald:

$$T_W^* = n\, r(\hat{\beta})^\top \left[R(\hat{\beta})(G_n^\top S^{-1} G_n)^{-1} R(\hat{\beta})^\top\right]^{-1} r(\hat{\beta})$$

   - DM:

$$T_{DM}^* = n\left[Q(\tilde{\beta}) - Q(\hat{\beta})\right]$$

4. **Compute the noncentrality parameter**

$$\lambda = \frac{1}{2}\, \mu_R^\top [R(\beta_0)(G_0^\top S^{-1} G_0)^{-1} R(\beta_0)^\top]^{-1} \mu_R.$$

5. **Determine the critical value**

$$\chi^2_{s,1-\alpha}.$$

6. **Compute power**

$$\text{Power} = P\left(\chi_s^2(\lambda) \geq \chi^2_{s,1-\alpha}\right).$$

Repeating these calculations over a grid of sample sizes yields power curves or tables for study planning. Additional practical implementation details, including instrument selection, construction of moment conditions, and extraction of variance components, are provided in Appendix B.

## 4. Simulation Study Framework

We conducted simulation studies to evaluate the finite-sample performance of the Wald-based and DM power methods under longitudinal GMM with time-dependent covariates. For each scenario, we specified a longitudinal data-generating mechanism, generated repeated datasets over a range of sample sizes and effect sizes, fit GMM using prespecified instruments, and computed both Wald and DM test statistics.

Empirical power was defined as the proportion of simulated datasets in which the null hypothesis was rejected at a given significance level. These empirical estimates were compared with the theoretical power obtained from the proposed Wald-based and DM formulas, allowing us to assess the accuracy of the asymptotic approximations across realistic sample sizes and feedback structures.

### 4.1. Example simulation design

We considered two settings adapted from Lai and Small (2007), each with a continuous outcome and a continuous time-dependent covariate of Type II or Type III. For each setting, we simulated data at three time points (t = 1, 2, 3) for nine sample sizes (n = 100, 200, 500, 1000, 2000, 3000, 4000, 5000, 10000), yielding 18 combinations of covariate type and sample size. These sample sizes were chosen based on prior work demonstrating that longitudinal GMM with time-dependent covariates typically requires larger samples for adequate power (Ramezani, 2017). In contrast to the smaller sample sizes examined in Ramezani (2017), we did not include n < 100 here because their behavior has already been extensively studied and shown to exhibit small-sample misalignment with asymptotic theory. The present study therefore focuses on the larger sample sizes needed to meaningfully evaluate asymptotic power and to assess the stability of the Wald and DM statistics in settings where asymptotic approximations are expected to hold.

The effect sizes used here are moderate and chosen to highlight differences in finite-sample stability between the Wald and DM statistics, consistent with prior findings on the sensitivity of distance-metric GMM criteria (Hall & Horowitz, 1996; Newey & Smith, 2004). For each combination, we generated 3600 datasets, consistent with previous recommendations on the number of replications (Ramezani, 2017).

Within-subject correlation and the number of time points were held fixed, following earlier GMM simulation studies (Liu & Liang, 1997, 1998; Lai & Small, 2007; Rochon, 1998). For each dataset, we fit the GMM model using the implementation of Ramezani and Wilson (2026), which employs the BFGS quasi-Newton algorithm to minimize the GMM quadratic form. This approach provides faster and more stable convergence than the derivative-free Nelder–Mead algorithm used in Ramezani (2017) and in earlier applied GMM work (Irimata et al., 2018; Lalonde et al., 2014). For comparison, we also estimated each dataset using the 2017 Nelder–Mead–based implementation. For each fit, we computed the Wald and DM statistics, estimated empirical power and type I error, and compared these to theoretical power based on the noncentrality parameter. All simulations and GMM estimations were conducted in R (R Core Team, 2015).

### 4.2. Data-generating mechanisms

**Setting 1 (Type II covariate).** Following Lai and Small (2007) and Ramezani and Wilson (2026), we generated data from

$$y_{it} = \gamma_0 + \gamma_1 x_{it} + \gamma_2 x_{i,t-1} + b_i + e_{it},$$

$$x_{it} = \rho x_{i,t-1} + \varepsilon_{it},$$

where $y_{it}$ and $x_{it}$ are the response and predictor for subject $i$ at time $t$. We set $\gamma_0 = 0, \gamma_1 = 1, \gamma_2 = 1, \rho = 0.5, b_i \sim N(0,4), e_{it} \sim N(0,1)$, and $\varepsilon_{it} \sim N(0,1)$, all mutually independent. The covariate process is stationary with $x_{i0} \sim N\{0, \sigma_\varepsilon^2/(1-\rho^2)\}$. This setting represents a Type II covariate where past covariate values affect both current covariates and responses.

**Setting 2 (Type III covariate).** In the second setting (Lai & Small, 2007) and Ramezani and Wilson (2026), we generated data from

$$y_{it} = \beta x_{it} + \kappa y_{i,t-1} + u_{it},$$

$$x_{it} = \gamma y_{i,t-1} + v_{it},$$

with $y_{it}$ and $x_{it}$ defined as above. We set $\beta = 0.5$, $\kappa = 0.3$, $\gamma = 0.5$, $u_{it} \sim N(0,1)$, and $v_{it} \sim N(0,1)$. This induces a feedback structure in which past responses influence both current covariates and responses. The joint process $(x_{it}, y_{it})$ is stationary with $y_{i0}$ generated from the corresponding stationary distribution. For each setting, we also simulated 3600 datasets under the null hypothesis (true effect size zero) to estimate type I error.

### 4.3. Results

Table 1 summarizes the empirical rejection rates and theoretical power for the Wald and DM statistics across all simulation settings using the BFGS-based GMM estimator. For Type II covariates, the Wald statistic closely matched its theoretical power even at smaller sample sizes, reflecting the relatively strong effect size in this setting. For Type III covariates, larger sample sizes were required for the Wald rejection rate to approach theoretical power, consistent with the stronger feedback structure.

Under the null hypothesis, the Wald statistic exhibited conservative behavior, with Type I error near zero across all settings. In contrast, the DM statistic showed substantial finite-sample inflation: its rejection rate was 1.000 for all sample sizes and both covariate types, and its Type I error ranged from 0.05 to 0.09. Although the gap between DM rejection rates and theoretical power narrowed as sample size increased, this occurred only because theoretical power approached one. This behavior is consistent with the well-documented sensitivity of the DM statistic to finite-sample variability in overidentified GMM settings (Hall & Horowitz, 1996; Newey & Smith, 2004).

Results were nearly identical when using the Nelder–Mead GMM estimator (Appendix C), confirming that the numerical optimization method did not materially affect the statistical performance of either test. However, the Nelder–Mead implementation required substantially more computation time and exhibited less stable convergence behavior, consistent with its derivative-free nature.

**Table 1:** Power analysis results

| | | | | Wald statistic | | DM statistic | |
|---|---|---|---|---|---|---|---|
| Time-Dependent Covariate | Sample Size | Theoretical Power | Theoretical Size (α = 0.05) | Rejection Rate | Type I Error Rate | Rejection Rate | Type I Error Rate |
| Type II | 100 | 0.9808 | 0.0671 | 0.9733 | 0.0003 | 1.0000 | 0.0914 |
| | 200 | 0.9818 | 0.0575 | 0.9800 | 0.0000 | 1.0000 | 0.0633 |
| | 500 | 0.9849 | 0.0527 | 0.9789 | 0.0000 | 1.0000 | 0.0536 |
| | 1000 | 0.9907 | 0.0514 | 0.9828 | 0.0000 | 1.0000 | 0.0575 |
| | 2000 | 0.9961 | 0.0506 | 0.9892 | 0.0000 | 1.0000 | 0.0522 |
| | 3000 | 0.9985 | 0.0504 | 0.9939 | 0.0000 | 1.0000 | 0.0558 |
| | 4000 | 0.9994 | 0.0503 | 0.9964 | 0.0000 | 1.0000 | 0.0497 |
| | 5000 | 0.9998 | 0.0503 | 0.9981 | 0.0000 | 1.0000 | 0.0475 |
| | 10000 | 1.0000 | 0.0501 | 1.0000 | 0.0000 | 1.0000 | 0.0519 |
| Type III | 100 | 0.3324 | 0.0519 | 0.1558 | 0.0000 | 1.0000 | 0.0781 |
| | 200 | 0.3606 | 0.0509 | 0.1964 | 0.0000 | 1.0000 | 0.0603 |
| | 500 | 0.4398 | 0.0503 | 0.3239 | 0.0000 | 1.0000 | 0.0544 |
| | 1000 | 0.5572 | 0.0502 | 0.5200 | 0.0000 | 1.0000 | 0.0558 |
| | 2000 | 0.7313 | 0.0501 | 0.7953 | 0.0000 | 1.0000 | 0.0511 |
| | 3000 | 0.8478 | 0.0501 | 0.9200 | 0.0000 | 1.0000 | 0.0475 |
| | 4000 | 0.9229 | 0.0500 | 0.9706 | 0.0000 | 1.0000 | 0.0500 |

|  | 5000 | 0.9587 | 0.0500 | 0.9919 | 0.0000 | 1.0000 | 0.0469 |
|---|---|---|---|---|---|---|---|
|  | 10000 | 0.9990 | 0.0500 | 1.0000 | 0.0000 | 1.0000 | 0.0483 |

**Note.** *Theoretical size equals α = 0.05; small deviations reflect numerical evaluation of the chi-square CDF.*

To assess distributional assumptions, we compared the empirical test statistics with the corresponding noncentral chi-square distribution using QQ-plots. For Type II covariates, agreement with theoretical quantiles improved steadily with sample size (Figure 1). For Type III covariates, substantially larger sample sizes were required for similar alignment (Figure 2). Complete QQ-plots for all sample sizes and both GMM algorithms are provided in Appendix D.

**Figure 1.** QQ-plots comparing empirical Wald statistics to the theoretical noncentral chi-square distribution (Type II)

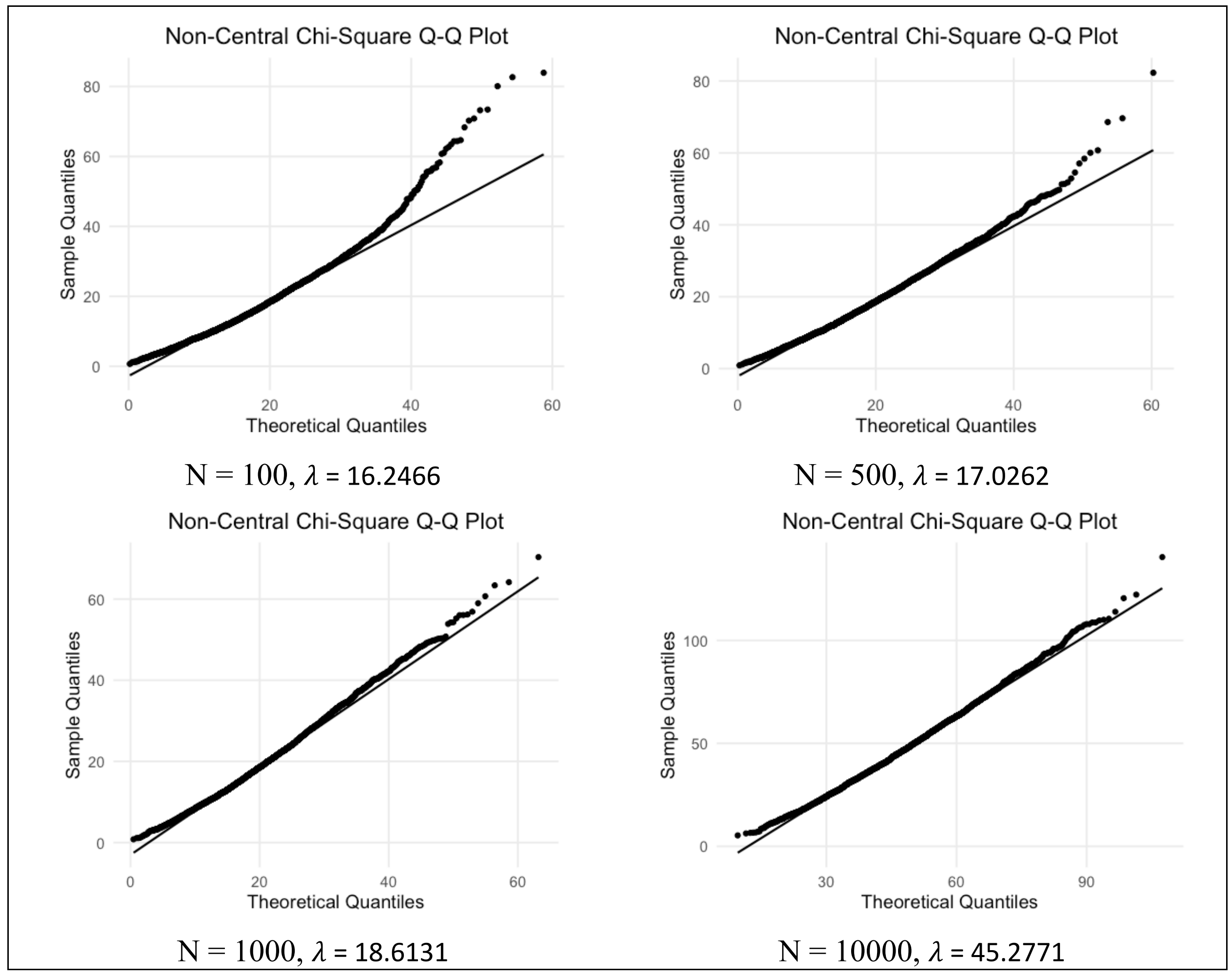

Figure 3 summarizes the theoretical power and empirical rejection rates for the Wald statistic across sample sizes for both Type II and Type III covariates. The curves show that theoretical power increases steadily with sample size in both settings, with Type II converging more rapidly. Empirical rejection rates closely track theoretical power, demonstrating the good finite-sample performance of the Wald statistic, particularly in the Type II setting.

**Figure 2.** QQ-plots comparing empirical Wald statistics to the theoretical noncentral chi-square distribution (Type III)

Non-Central Chi-Square Q-Q Plot

Sample Quantiles

Theoretical Quantiles

N = 100, $\lambda$ = 2.3287

Non-Central Chi-Square Q-Q Plot

Sample Quantiles

Theoretical Quantiles

N = 500, $\lambda$ = 3.2701

Non-Central Chi-Square Q-Q Plot

Sample Quantiles

Theoretical Quantiles

N = 1000, $\lambda$ = 4.4258

Non-Central Chi-Square Q-Q Plot

Sample Quantiles

Theoretical Quantiles

N = 10000, $\lambda$ = 25.385

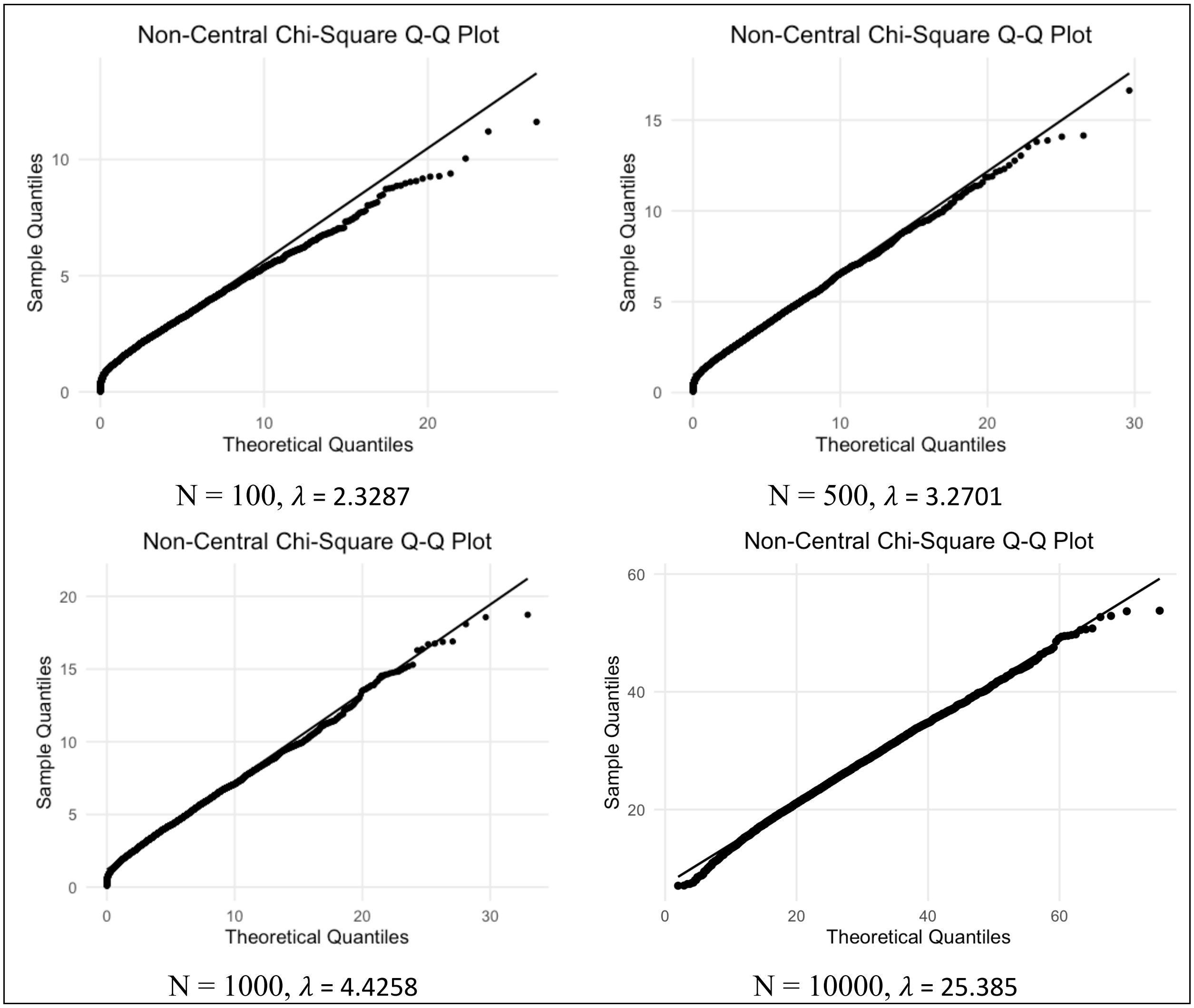

**Figure 3.** Wald statistic power and rejection rate across sample sizes

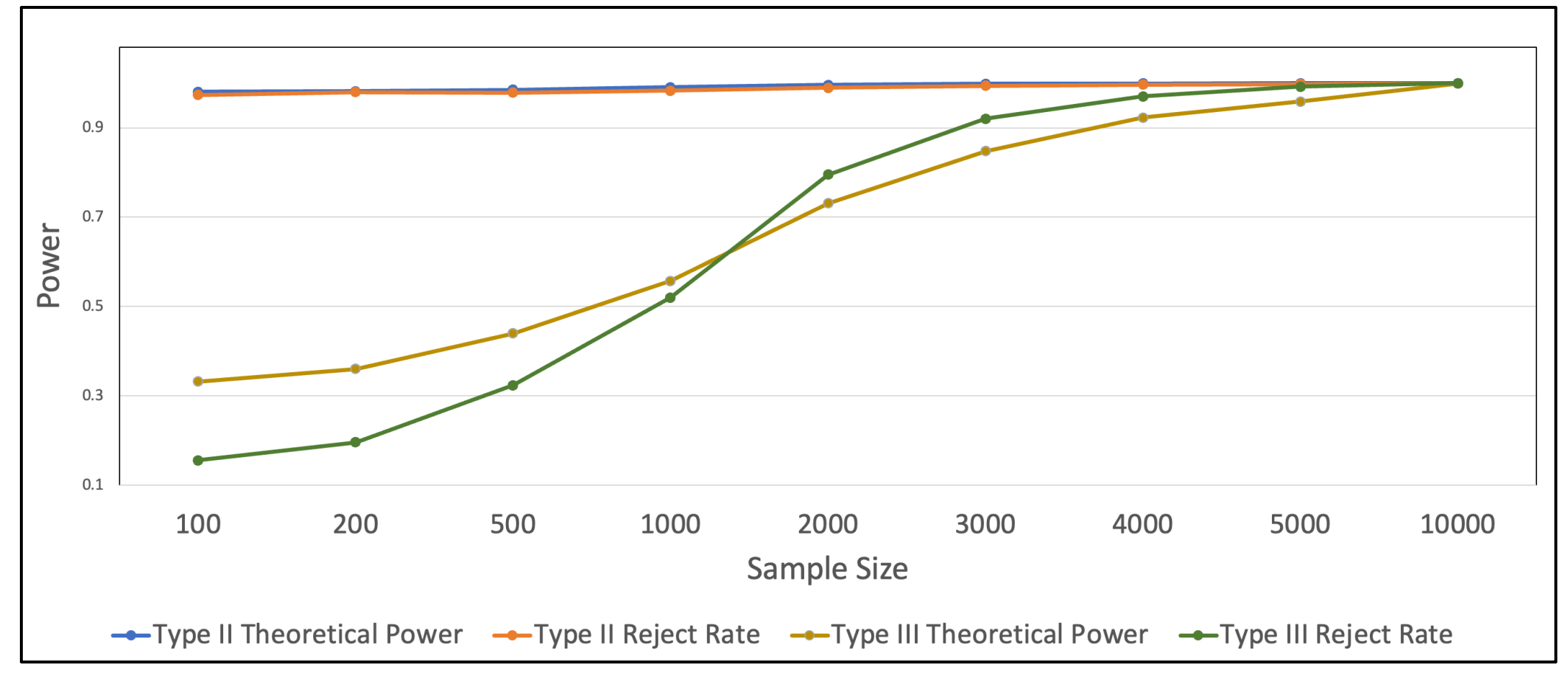

### 4.4. Interpretations

The Wald statistic demonstrated good finite-sample performance in both settings. For Type II covariates, empirical rejection rates were close to theoretical power even at $n = 100$. For Type III covariates, alignment occurred only at larger sample sizes (approximately $n \approx 2000$), reflecting the stronger feedback and slower convergence of the noncentral chi-square approximation.

For the DM statistic, empirical rejection rates were uniformly equal to 1.000 across all sample sizes, indicating severe finite-sample inflation. This behavior is not an artifact of the simulation design but a well-documented consequence of the DM statistic's instability in overidentified GMM settings (Hall & Horowitz, 1996; Newey & Smith, 2004). Even when theoretical power is moderate, the DM statistic can behave as if the alternative is overwhelmingly strong, producing rejection rates near one. This contrast with the Wald statistic is precisely the phenomenon we aim to illustrate: Wald remains calibrated, whereas DM does not.

The difference between empirical and theoretical power diminished only as theoretical power approached one. Alignment occurred at $n \approx 1000$ for Type II covariates and at $n \approx 10000$ for Type III covariates. This pattern is consistent with the known sensitivity of the DM statistic to sampling variability in overidentified GMM models, particularly when moment conditions are correlated or when feedback is strong.

The finite-sample inflation of the DM statistic arises because the GMM minimand is highly sensitive to sampling variability in overidentified settings, causing even small deviations in moment conditions to produce large changes in the objective function.

## 5. Application to the Osteoarthritis Initiative

### 5.1. Study overview

To illustrate the implementation of the proposed power estimation methods, we applied them to data from the OAI, a large multi-center longitudinal cohort study of adults aged 45 and older. The analytic sample includes repeated measurements of knee pain and covariates over three follow-up visits. The primary outcome is the continuous Western Ontario and McMaster Universities (WOMAC) disability score, which quantifies pain, stiffness, and physical function in knee osteoarthritis. We used the average of the left- and right-knee WOMAC scores as the continuous response, reflecting overall knee symptom severity in the presence of time-dependent covariates (Bellamy et al., 1988).

### 5.2. Model specification

We specified a marginal mean model of the form

$$\mu_{it} = g^{-1}(X_{it}^{\top}\beta),$$

using the identity link. The covariate vector included sex, age, BMI, and two follow-up time indicators. Because BMI may change in response to prior pain, it serves as an internal time-dependent covariate, making GMM an appropriate estimation framework.

### 5.3. Instrument selection and moment conditions

Following Lai and Small (2007), instruments were constructed from lagged outcomes, lagged covariates, and baseline characteristics. These instruments satisfy the requirement of being correlated with the endogenous covariates but uncorrelated with contemporaneous errors. The stacked moment conditions take the form

$$E[Z_{it}(Y_{it} - \mu_{it}(\beta))] = 0.$$

### 5.4. GMM estimation and variance extraction

The GMM estimator was obtained by minimizing the quadratic form

$$\bar{m}_n(\beta)^{\top} W_n^{-1} \bar{m}_n(\beta),$$

Table 2 summarizes the estimated coefficients. The asymptotic variance of the BMI coefficient was extracted and used to compute the noncentrality parameters for the Wald-based and distance-metric power calculations.

**Table 2.** GMM Model Summary

| Parameters | Intercept | Sex | Age | BMI | $t_2$ | $t_3$ |
|---|---|---|---|---|---|---|
| Coefficients | -1.48913 | 0.20018 | 0.00255 | 0.72824 | -0.11059 | -0.10393 |

### 5.5. Power estimation

To estimate power for detecting an association between BMI and WOMAC pain, we specified a set of clinically meaningful effect sizes for the BMI coefficient. For each effect size, the noncentrality parameter was computed as

$$\lambda(n) = n \cdot \frac{(\beta_{j1} - \beta_{j0})^2}{\sigma_j^2},$$

where $\sigma_j^2$ is the asymptotic variance of the BMI coefficient. We report power using both subsample-based GMM estimates and population-level GMM estimates to distinguish between power calculations that rely on variance estimates available in a smaller pilot sample versus those based on the full cohort. Table 3 summarizes the resulting theoretical power values across sample sizes. The former reflects what investigators would compute during study planning, while the latter represents the theoretical power under the true population parameters.

**Table 3.** Theoretical Power for BMI Effect Across Sample Sizes

| **Sample Size** | **λ (Subsample GMM)** | **Power (Subsample)** | **λ (Population GMM)** | **Power (Population)** |
|---|---|---|---|---|
| 25 | 5.756 | 0.670 | 9.033 | 0.852 |
| 50 | 6.070 | 0.693 | 10.097 | 0.888 |
| 100 | 7.296 | 0.771 | 10.496 | 0.900 |
| 200 | 7.479 | 0.781 | 10.821 | 0.908 |

These results illustrate how the proposed framework can be used to evaluate the detectability of clinically meaningful BMI effects in longitudinal studies. Power estimates based on subsample-level GMM variances reflect what investigators would compute during study planning, while population-level estimates represent the theoretical power under the true variance structure. Together, these calculations demonstrate how GMM-based power analysis can guide study design decisions when internal time-dependent covariates are present.

### 5.6 Interpretation

The OAI example demonstrates that GMM-based power estimation is feasible in real longitudinal data with internal time-dependent covariates. The extracted asymptotic variance of the BMI coefficient provides the necessary inputs for computing noncentrality parameters, the resulting power values show the expected increases in power with sample size. Because the OAI analysis relies on asymptotic variance estimates rather than empirical test statistics, the finite-sample instability of the DM statistic observed in the simulations does not affect the power calculations presented here. These results illustrate how the proposed methods can support study planning in settings where covariates evolve in response to prior outcomes.

## 6. Discussion

The methods developed in this paper address a longstanding gap in the literature on power estimation for longitudinal studies with internal time-dependent covariates. Existing approaches based on generalized estimating equations are not valid when covariates depend on past outcomes, as the working-correlation becomes misspecified and estimators may be biased. GMM provides a natural alternative in this setting, because it relies on valid moment conditions and avoids full joint distributional assumptions. Despite the advantages of GMM for estimation, no practical framework for power analysis under GMM has been available. This paper fills that gap by developing two complementary power estimation methods—one based on the Wald statistic and one based on the distance-metric statistic—and by demonstrating their implementation using both theoretical derivations and applied examples. This study also extends earlier work (Ramezani, 2017) by examining the larger sample sizes required for asymptotic power calculations, addressing the small-sample limitations previously identified.

The Wald-based method provides a direct and interpretable approach to power estimation, relying on the asymptotic normality of the GMM estimator and the variance of the parameter of interest. The distance-metric method offers a more global assessment of the discrepancy between the null and alternative hypotheses by incorporating the full set of moment conditions. Together, these methods provide investigators with flexible tools for planning longitudinal studies in which covariates evolve in response to past outcomes. The OAI example illustrates how these methods can be applied in practice, while the simulation framework evaluates their finite-sample behavior.

The simulation results highlight several important differences in the finite-sample behavior of the two statistics. First, the Wald statistic aligns with theoretical power at moderate sample sizes in Type II covariate settings ($n \approx 100$) and requires larger samples in Type III settings ($n \approx 2000$). This pattern reflects the stronger feedback and slower convergence induced by internal time-dependent covariates in the Type III design.

In contrast, the DM statistic exhibited uniformly inflated rejection rates, with values equal to 1.000 across all sample sizes and both covariate types. Apparent "alignment" occurred only when theoretical power approached one, a consequence of ceiling effects rather than accurate calibration. This behavior reflects well-documented finite-sample distortions of difference-in-objective-function tests in overidentified GMM models. These tests are sensitive to the correlation structure of the moment conditions and to sampling variability in the estimated weighting matrix, leading to liberal behavior even when the Wald statistic remains well controlled (Hall & Horowitz, 1996; Newey & Smith, 2004). These effects are amplified in Type III settings due to stronger feedback and greater moment correlation.

Additional diagnostics support these conclusions. Both BFGS and Nelder–Mead optimization algorithms produced nearly identical estimates and test statistics, confirming that the observed patterns are not artifacts of the optimization routine. QQ plots further showed faster convergence for Type II covariates and slower convergence for Type III covariates, consistent with theoretical expectations.

Taken together, these findings clarify the practical implications of using Wald-based versus DM-based power calculations in longitudinal GMM settings. The Wald statistic provides stable and interpretable power estimates in moderate sample sizes, whereas the DM statistic should be used with caution due to its finite-sample inflation. The proposed framework therefore offers a principled and implementable approach for planning longitudinal studies with internal time-dependent covariates, while also highlighting the importance of understanding the finite-sample behavior of GMM test statistics.

## 7. Limitations

Several limitations of the proposed methods warrant consideration. First, the accuracy of the power estimates depends on the quality of the asymptotic variance estimates obtained from the GMM model. In small samples or in models with many moment conditions, these variance estimates may be unstable, affecting the precision of the resulting power calculations. Second, the choice of instruments plays a critical role in GMM performance. Weak or poorly chosen instruments can inflate variances and reduce power, even when the moment conditions are theoretically valid. Third, the methods assume correct specification of both the marginal mean model and the moment conditions; misspecification of either component may lead to biased estimates and inaccurate power calculations.

A further limitation is that the theoretical power formulas rely on asymptotic approximations. Although these approximations perform well in moderate to large samples, their accuracy in small samples may be limited. This is particularly relevant for the distance-metric statistic, whose finite-sample distribution is sensitive to sampling variability in the weighting matrix and to the correlation structure of the moment conditions. As demonstrated in our simulations, these properties can lead to inflated rejection rates in small to moderate samples. Simulation studies can help assess the extent of this limitation, but they require additional computational effort.

## 8. Future Directions

Several avenues for future research emerge from this work. One important direction is the development of power estimation methods for nonlinear GMM models, including logistic and Poisson marginal models with time-dependent covariates. Extending the framework to accommodate non-Gaussian outcomes or nonlinear marginal models represents a natural next step, as the current work focuses on continuous outcomes and linear specifications. Another promising area is the incorporation of model selection procedures into the power estimation framework, particularly in settings with many potential instruments or moment conditions. Adaptive or data-driven instrument selection methods may improve both estimation efficiency and power.

Future work may also explore the use of bootstrap or resampling methods to improve the accuracy of power estimates in small samples, providing more reliable variance estimates and reducing reliance on asymptotic approximations. Additionally, extending the framework to accommodate missing data, dropout, or irregular observation times would increase its applicability to real-world longitudinal studies. Finally, the development of user-friendly software tools implementing the proposed methods would facilitate their adoption by applied researchers and practitioners.

## 9. Conclusion

This paper develops a practical and theoretically grounded framework for power estimation in longitudinal studies with internal time-dependent covariates using the Generalized Method of Moments. By deriving power formulas based on both the Wald statistic and the distance-metric statistic, and by demonstrating their implementation through the OAI example, the paper provides investigators with tools that were previously unavailable for this class of models. The methods address a critical gap in the literature and offer a flexible, robust approach to planning longitudinal studies in which covariates evolve over time. As longitudinal data continue to play a central role in biomedical, behavioral, and social science research, the ability to conduct valid and efficient power analyses under GMM will be increasingly important.

In particular, the results highlight the practical advantages of Wald-based power calculations in moderate sample sizes and the challenges posed by finite-sample distortions in distance-metric statistics. These insights provide guidance for applied researchers and point toward future methodological refinements that can further strengthen the use of GMM in longitudinal study design.

**Acknowledgments.**

The authors thank Dr. Trent Lalonde for early conceptual discussions and guidance during the dissertation phase of this work.

## APPENDIX A. Proofs of Theorem 1 and Theorem 2

### A.1 Theorem A.1 (Quadratic Forms in Normal Variables).

According to this theorem (Ravishanker & Dey, 2002), if $\boldsymbol{Y}$, a random vector, follows a normal distribution of $N(\boldsymbol{\mu}, \boldsymbol{\Sigma})$ where $\boldsymbol{\Sigma}$ is a full rank positive definite matrix, $\boldsymbol{A}$ is a symmetric matrix with $rank(\boldsymbol{A}) = m$; then, $\boldsymbol{Y}^T\boldsymbol{A}\boldsymbol{Y} \sim \chi^2\left(m, \frac{\boldsymbol{\mu}^T\boldsymbol{A}\boldsymbol{\mu}}{2}\right)$ if any one of the following three conditions are met:

1. $\boldsymbol{A\Sigma}$ is an idempotent matrix of rank $m$.
2. $\boldsymbol{\Sigma A}$ is an idempotent matrix of rank $m$.
3. $\boldsymbol{\Sigma}$ is a g-inverse of $\boldsymbol{A}$ with $rank(\boldsymbol{A}) = m$.

This theorem is used to derive the asymptotic distribution of the Wald statistic in Sections A.2.1 (special case) and A.2.2 (general case).

### A.2 Proof of Theorem 1 (Asymptotic Distribution of the Wald Statistic)

#### A.2.1 Special Case: $\mathbf{h_0} = \mathbf{0}$

Consider the Wald statistic specified in Section 3.4. It can be written as below

$$T_W^* = n\left(\boldsymbol{H}\widehat{\boldsymbol{\beta}} - \boldsymbol{h}_0\right)^T \left[R\left(\widehat{\boldsymbol{\beta}}\right)\left(\boldsymbol{G}_n\left(\widehat{\boldsymbol{\beta}}\right)^T\boldsymbol{S}^{-1}\boldsymbol{G}_n\left(\widehat{\boldsymbol{\beta}}\right)\right)^{-1} R\left(\widehat{\boldsymbol{\beta}}\right)^T\right]^{-1} \left(\boldsymbol{H}\widehat{\boldsymbol{\beta}} - \boldsymbol{h}_0\right),$$

in which $R(\boldsymbol{\beta}) = \frac{\partial r(\boldsymbol{\beta})}{\partial \boldsymbol{\beta}'}$ will be $\mathbf{H}$ after taking the derivative of $r(\boldsymbol{\beta})$. For the sake of simplicity, let's write $\boldsymbol{G}_n\left(\widehat{\boldsymbol{\beta}}\right) = \boldsymbol{G}$. Now the Wald Statistic equation can be written as

$$T_W^* = n\left(\boldsymbol{H}\widehat{\boldsymbol{\beta}} - \boldsymbol{h}_0\right)^T [\boldsymbol{H}(\boldsymbol{G}^T\boldsymbol{S}^{-1}\boldsymbol{G})^{-1}\boldsymbol{H}^T]^{-1}\left(\boldsymbol{H}\widehat{\boldsymbol{\beta}} - \boldsymbol{h}_0\right).$$

Defining $[\boldsymbol{H}(\boldsymbol{G}^T\boldsymbol{S}^{-1}\boldsymbol{G})^{-1}\boldsymbol{H}^T]^{-1} = \boldsymbol{B},$ $T_W^*$ can be simplified as

$$\begin{aligned} T_W^* &= n\left(\widehat{\boldsymbol{\beta}}^T\boldsymbol{H}^T - \boldsymbol{h}_0^T\right)\boldsymbol{B}\left(\boldsymbol{H}\widehat{\boldsymbol{\beta}} - \boldsymbol{h}_0\right) \\ &= n\left[\widehat{\boldsymbol{\beta}}^T\boldsymbol{H}^T\boldsymbol{B}\boldsymbol{H}\widehat{\boldsymbol{\beta}} - \widehat{\boldsymbol{\beta}}^T\boldsymbol{H}^T\boldsymbol{B}\boldsymbol{h}_0 - \boldsymbol{h}_0^T\boldsymbol{B}\boldsymbol{H}\widehat{\boldsymbol{\beta}} + \boldsymbol{h}_0^T\boldsymbol{B}\boldsymbol{h}_0\right]. \end{aligned} \quad \text{(A.1)}$$

Under the common special case that $\boldsymbol{h}_0 = \mathbf{0}$ and by substituting $\boldsymbol{B},$ Equation A.1 can simply be written as

$$\begin{aligned} T_W^* &= n\left[\widehat{\boldsymbol{\beta}}^T\boldsymbol{H}^T\boldsymbol{B}\boldsymbol{H}\widehat{\boldsymbol{\beta}}\right] = n\left[\widehat{\boldsymbol{\beta}}^T\boldsymbol{H}^T[\boldsymbol{H}(\boldsymbol{G}^T\boldsymbol{S}^{-1}\boldsymbol{G})^{-1}\boldsymbol{H}^T]^{-1}\boldsymbol{H}\widehat{\boldsymbol{\beta}}\right] \\ &= \left(\sqrt{n}\widehat{\boldsymbol{\beta}}^T\right)\boldsymbol{H}^T[\boldsymbol{H}(\boldsymbol{G}^T\boldsymbol{S}^{-1}\boldsymbol{G})^{-1}\boldsymbol{H}^T]^{-1}\boldsymbol{H}\left(\sqrt{n}\widehat{\boldsymbol{\beta}}\right). \end{aligned} \quad \text{(A.2)}$$

Using Theorem A.1, assume

$$\boldsymbol{A} = \boldsymbol{H}^T[\boldsymbol{H}(\boldsymbol{G}^T\boldsymbol{S}^{-1}\boldsymbol{G})^{-1}\boldsymbol{H}^T]^{-1}\boldsymbol{H},$$

and

$$\boldsymbol{Y} = \sqrt{n}\widehat{\boldsymbol{\beta}}.$$

Because $\widehat{\boldsymbol{\beta}}$ is asymptotically normal, $\sqrt{n}\widehat{\boldsymbol{\beta}} \dot{\sim} N(\sqrt{n}\boldsymbol{\beta}, \boldsymbol{\Sigma})$, where $\boldsymbol{\Sigma} = (\boldsymbol{G}^T\boldsymbol{S}^{-1}\boldsymbol{G})^{-1}$. It is shown below that $\boldsymbol{A\Sigma}$ is an idempotent matrix meaning that $(\boldsymbol{A\Sigma})(\boldsymbol{A\Sigma}) = \boldsymbol{A\Sigma}$. Substituting $\boldsymbol{A}$ and $\boldsymbol{\Sigma}$,

$$\begin{aligned}&(\boldsymbol{A\Sigma})(\boldsymbol{A\Sigma}) \\ &= \{\boldsymbol{H}^T[\boldsymbol{H}(\boldsymbol{G}^T\boldsymbol{S}^{-1}\boldsymbol{G})^{-1}\boldsymbol{H}^T]^{-1}\boldsymbol{H}(\boldsymbol{G}^T\boldsymbol{S}^{-1}\boldsymbol{G})^{-1}\}\{\boldsymbol{H}^T[\boldsymbol{H}(\boldsymbol{G}^T\boldsymbol{S}^{-1}\boldsymbol{G})^{-1}\boldsymbol{H}^T]^{-1}\boldsymbol{H}(\boldsymbol{G}^T\boldsymbol{S}^{-1}\boldsymbol{G})^{-1}\}.\end{aligned}$$

Due to the fact that $\boldsymbol{H}(\boldsymbol{G}^T\boldsymbol{S}^{-1}\boldsymbol{G})^{-1}\,\boldsymbol{H}^T[\boldsymbol{H}(\boldsymbol{G}^T\boldsymbol{S}^{-1}\boldsymbol{G})^{-1}\boldsymbol{H}^T]^{-1} = \boldsymbol{I}$, then

$$\begin{aligned}(\boldsymbol{A\Sigma})(\boldsymbol{A\Sigma}) &= \boldsymbol{H}^T[\boldsymbol{H}(\boldsymbol{G}^T\boldsymbol{S}^{-1}\boldsymbol{G})^{-1}\boldsymbol{H}^T]^{-1}\boldsymbol{I}\boldsymbol{H}(\boldsymbol{G}^T\boldsymbol{S}^{-1}\boldsymbol{G})^{-1} \\ &= \boldsymbol{H}^T[\boldsymbol{H}(\boldsymbol{G}^T\boldsymbol{S}^{-1}\boldsymbol{G})^{-1}\boldsymbol{H}^T]^{-1}\boldsymbol{H}(\boldsymbol{G}^T\boldsymbol{S}^{-1}\boldsymbol{G})^{-1} = \boldsymbol{A\Sigma}.\end{aligned}$$

This proves that $\boldsymbol{A\Sigma}$ is idempotent, which is the one condition that needs to be met in order to conclude that the quadratic form Equation A.2 is distributed as a chi-square. So, substituting $\boldsymbol{Y} = \sqrt{n}\widehat{\boldsymbol{\beta}}$ and $\boldsymbol{\mu} = \sqrt{n}\boldsymbol{\beta}$ in $\boldsymbol{Y}^T\boldsymbol{A}\boldsymbol{Y} \sim \chi^2\left(m, \frac{\boldsymbol{\mu}^T\boldsymbol{A}\boldsymbol{\mu}}{2}\right)$, we can say

$$\boldsymbol{Y}^T\boldsymbol{A}\boldsymbol{Y} = \left(\sqrt{n}\widehat{\boldsymbol{\beta}}\right)^T\boldsymbol{H}^T[\boldsymbol{H}(\boldsymbol{G}^T\boldsymbol{S}^{-1}\boldsymbol{G})^{-1}\boldsymbol{H}^T]^{-1}\boldsymbol{H}\sqrt{n}\widehat{\boldsymbol{\beta}} = T_W^* \sim \chi^2(s, \lambda),$$

where the non-centrality parameter under the null hypothesis is defined as

$$\begin{aligned}\lambda = \frac{\boldsymbol{\mu}^T\boldsymbol{A}\boldsymbol{\mu}}{2} &= \frac{1}{2}\left(\sqrt{n}\boldsymbol{\beta_0}\right)^T\boldsymbol{H}^T[\boldsymbol{H}(\boldsymbol{G}_0^T\boldsymbol{S}^{-1}\boldsymbol{G_0})^{-1}\boldsymbol{H}^T]^{-1}\boldsymbol{H}\left(\sqrt{n}\boldsymbol{\beta_0}\right) \\ &= \frac{1}{2}\boldsymbol{\mu}_R^T[R(\boldsymbol{\beta}_0)(\boldsymbol{G}_0^T\boldsymbol{S}^{-1}\boldsymbol{G_0})^{-1}R(\boldsymbol{\beta}_0)^T]^{-1}\boldsymbol{\mu}_R,\end{aligned}$$

defining $\boldsymbol{\mu}_R = \sqrt{n}\boldsymbol{\beta_0}$. □

**A.2.2 General Case: $\mathbf{h_0 \neq 0}$**

Knowing that $r(\widehat{\boldsymbol{\beta}}) = \boldsymbol{H}\widehat{\boldsymbol{\beta}} - \boldsymbol{h}_0 \sim N(\boldsymbol{H\beta} - \boldsymbol{h}_0, \boldsymbol{H\Sigma H}^T)$, consider the Wald statistic

$$\begin{aligned}T_W^* &= n\left(r(\widehat{\boldsymbol{\beta}})\right)^T\left[R(\widehat{\boldsymbol{\beta}})\left(\boldsymbol{G}_n(\widehat{\boldsymbol{\beta}})^T\boldsymbol{S}^{-1}\boldsymbol{G}_n(\widehat{\boldsymbol{\beta}})\right)^{-1}R(\widehat{\boldsymbol{\beta}})^T\right]^{-1}\left(r(\widehat{\boldsymbol{\beta}})\right) \\ &= n\left(r(\widehat{\boldsymbol{\beta}})\right)^T[\boldsymbol{H}(\boldsymbol{G}^T\boldsymbol{S}^{-1}\boldsymbol{G})^{-1}\boldsymbol{H}^T]^{-1}\left(r(\widehat{\boldsymbol{\beta}})\right) \\ &= \left(\sqrt{n}r(\widehat{\boldsymbol{\beta}})^T\right)[\boldsymbol{H}(\boldsymbol{G}^T\boldsymbol{S}^{-1}\boldsymbol{G})^{-1}\boldsymbol{H}^T]^{-1}\left(\sqrt{n}r(\widehat{\boldsymbol{\beta}})\right).\end{aligned}$$

This time $\boldsymbol{A}$ and $\boldsymbol{Y}$ are different from before; calling them $\boldsymbol{A}^*$ and $\boldsymbol{Y}^*$, which are

$$\boldsymbol{A}^* = [\boldsymbol{H}(\boldsymbol{G}^T\boldsymbol{S}^{-1}\boldsymbol{G})^{-1}\boldsymbol{H}^T]^{-1},$$

and

$$\boldsymbol{Y}^* = \sqrt{n}r(\widehat{\boldsymbol{\beta}}),$$

which still follows a normal distribution but with a different mean and variance. To find the mean and variance of $\boldsymbol{Y}^*$, consider

$$\sqrt{n}\widehat{\boldsymbol{\beta}} \dot{\sim} N\left(\sqrt{n}\boldsymbol{\beta}, \boldsymbol{\Sigma}\right) \rightarrow \sqrt{n}\boldsymbol{H}\widehat{\boldsymbol{\beta}} \dot{\sim} N\left(\sqrt{n}\boldsymbol{H}\boldsymbol{\beta}, \boldsymbol{H}\boldsymbol{\Sigma}\boldsymbol{H}^{\boldsymbol{T}}\right)$$

$$\rightarrow \sqrt{n}\left(\boldsymbol{H}\widehat{\boldsymbol{\beta}} - \boldsymbol{h}_0\right) \dot{\sim} N\left(\sqrt{n}(\boldsymbol{H}\boldsymbol{\beta} - \boldsymbol{h}_0), \boldsymbol{H}\boldsymbol{\Sigma}\boldsymbol{H}^{\boldsymbol{T}}\right),$$

so, $\boldsymbol{Y}^* \sim N\left(\sqrt{n}(\boldsymbol{H}\boldsymbol{\beta} - \boldsymbol{h}_0), \boldsymbol{H}\boldsymbol{\Sigma}\boldsymbol{H}^{\boldsymbol{T}}\right)$.

Knowing that the new variance covariance matrix of $\boldsymbol{Y}^*$ is

$$\boldsymbol{\Sigma}^* = \boldsymbol{H}\boldsymbol{\Sigma}\boldsymbol{H}^{\boldsymbol{T}} = \boldsymbol{H}(\boldsymbol{G}^T\boldsymbol{S}^{-1}\boldsymbol{G})^{-1}\boldsymbol{H}^{\boldsymbol{T}},$$

it can be shown that $\boldsymbol{A}^*\boldsymbol{\Sigma}^*$ is an identity matrix, hence an idempotent one,

$$(\boldsymbol{A}^*\boldsymbol{\Sigma}^*)(\boldsymbol{A}^*\boldsymbol{\Sigma}^*) =$$
$$= \{[\boldsymbol{H}(\boldsymbol{G}^T\boldsymbol{S}^{-1}\boldsymbol{G})^{-1}\boldsymbol{H}^T]^{-1}\boldsymbol{H}(\boldsymbol{G}^T\boldsymbol{S}^{-1}\boldsymbol{G})^{-1}\boldsymbol{H}^T\}\{[\boldsymbol{H}(\boldsymbol{G}^T\boldsymbol{S}^{-1}\boldsymbol{G})^{-1}\boldsymbol{H}^T]^{-1}\boldsymbol{H}(\boldsymbol{G}^T\boldsymbol{S}^{-1}\boldsymbol{G})^{-1}\boldsymbol{H}^T\} = I^2 = I$$
$$= \boldsymbol{A}^*\boldsymbol{\Sigma}^*.$$

Therefore,

$$\boldsymbol{Y}^{*T}\boldsymbol{A}^*\boldsymbol{Y}^* = \left(\sqrt{n}r\left(\widehat{\boldsymbol{\beta}}\right)^T\right)[\boldsymbol{H}(\boldsymbol{G}^T\boldsymbol{S}^{-1}\boldsymbol{G})^{-1}\boldsymbol{H}^T]^{-1}\left(\sqrt{n}r\left(\widehat{\boldsymbol{\beta}}\right)\right) \sim \chi^2(s, \lambda),$$

where the non-centrality parameter is as below under the null hypothesis

$$\lambda^* = \frac{\boldsymbol{\mu}^{*T}\boldsymbol{A}^*\boldsymbol{\mu}^*}{2} = \frac{1}{2}\left(\sqrt{n}(\boldsymbol{H}\boldsymbol{\beta}_{\boldsymbol{0}} - \boldsymbol{h}_0)\right)^T[\boldsymbol{H}(\boldsymbol{G}_0^T\boldsymbol{S}^{-1}\boldsymbol{G}_{\boldsymbol{0}})^{-1}\boldsymbol{H}^T]^{-1}\left(\sqrt{n}(\boldsymbol{H}\boldsymbol{\beta}_{\boldsymbol{0}} - \boldsymbol{h}_0)\right)$$
$$= \frac{1}{2}\boldsymbol{\mu}^{*T}_R[R(\boldsymbol{\beta}_0)(\boldsymbol{G}_0^T\boldsymbol{S}^{-1}\boldsymbol{G}_{\boldsymbol{0}})^{-1}R(\boldsymbol{\beta}_0)^T]^{-1}\boldsymbol{\mu}^*_R,$$

defining $\boldsymbol{\mu}^*_R = \sqrt{n}(\boldsymbol{H}\boldsymbol{\beta}_{\boldsymbol{0}} - \boldsymbol{h}_0)$. □

## Appendix B. Practical implementation Details

In practice, both power methods require the same underlying components. First, the marginal model

$$\mu_{it} = g^{-1}(X_{it}^{\top}\boldsymbol{\beta})$$

must be specified. Second, valid instruments must be chosen; following Lai and Small (2007), lagged outcomes and covariates, as well as baseline covariates, are natural candidates. Third, the moment conditions are constructed, for example as

$$\boldsymbol{m}_i(\boldsymbol{\beta}) = \boldsymbol{Z}_i^{\top}(\boldsymbol{Y}_i - \boldsymbol{\mu}_i(\boldsymbol{\beta})),$$

where $\boldsymbol{Z}_i$ stacks the chosen instruments and $\boldsymbol{Y}_i$ and $\boldsymbol{\mu}_i(\boldsymbol{\beta})$ stack the outcomes and marginal means over time for subject $i$. Fourth, a GMM model is estimated using pilot or historical data to obtain $\widehat{\boldsymbol{\beta}}$ and an estimate $\widehat{\boldsymbol{V}}$ of the asymptotic covariance matrix. Fifth, the effect size $\beta_{j1} - \beta_{j0}$ is defined for the parameter of interest. Sixth, the asymptotic variance component $\sigma_j^2$ is extracted from $\widehat{\boldsymbol{V}}$. Finally, power is computed using either the Wald-based formulation, which relies on the noncentral chi-square distribution with one degree of freedom, or the distance-metric formulation, which uses the noncentral chi-square distribution with degrees of freedom equal to the number of overidentifying restrictions. Repeating these calculations over a grid of sample sizes yields power curves or tables suitable for study planning.

**Appendix C. Nelder–Mead GMM Results**

This appendix reports results obtained using the Nelder–Mead–based GMM estimator, which corresponds to the implementation used in earlier GMM literature. These results serve as a comparison to the BFGS quasi-Newton implementation. As shown below, the qualitative patterns for both the Wald and DM statistics are consistent across the two optimization methods.

**Table C1:** Wald statistic results using the Nelder–Mead GMM estimator

| Time-Dependent Covariate | Sample Size | Rejection Rate | Theoretical Power | Type I Error Rate | Theoretical Size ($\alpha$ = 0.05) |
|---|---|---|---|---|---|
| Type II | 100 | 0.9733 | 0.9808 | 0.0008 | 0.0685 |
| | 200 | 0.9806 | 0.9818 | 0.0000 | 0.0577 |
| | 500 | 0.9797 | 0.9849 | 0.0000 | 0.0527 |
| | 1000 | 0.9828 | 0.9907 | 0.0000 | 0.0514 |
| | 2000 | 0.9892 | 0.9961 | 0.0000 | 0.0506 |
| | 3000 | 0.9939 | 0.9985 | 0.0000 | 0.0504 |
| | 4000 | 0.9964 | 0.9994 | 0.0000 | 0.0503 |
| | 5000 | 0.9981 | 0.9998 | 0.0000 | 0.0503 |
| | 10000 | 1.0000 | 1.0000 | 0.0000 | 0.0501 |
| Type III | 100 | 0.1611 | 0.3324 | 0.0000 | 0.0521 |
| | 200 | 0.1986 | 0.3606 | 0.0000 | 0.0509 |
| | 500 | 0.3247 | 0.4398 | 0.0000 | 0.0503 |
| | 1000 | 0.5211 | 0.5572 | 0.0000 | 0.0502 |
| | 2000 | 0.7956 | 0.7313 | 0.0000 | 0.0501 |
| | 3000 | 0.9200 | 0.8478 | 0.0000 | 0.0501 |
| | 4000 | 0.9706 | 0.9229 | 0.0000 | 0.0500 |
| | 5000 | 0.9919 | 0.9587 | 0.0000 | 0.0500 |
| | 10000 | 1.0000 | 0.9990 | 0.0000 | 0.0500 |

**Table C2.** DM statistic results using the Nelder–Mead GMM estimator.

| Time-Dependent Covariate | Sample Size | Rejection Rate | Theoretical Power | Type I Error Rate | Theoretical Size ($\alpha$ = 0.05) |
|---|---|---|---|---|---|
| Type II | 100 | 1.0000 | 0.9808 | 0.0575 | 0.0685 |
| | 200 | 1.0000 | 0.9818 | 0.0511 | 0.0577 |
| | 500 | 1.0000 | 0.9849 | 0.0464 | 0.0527 |
| | 1000 | 1.0000 | 0.9907 | 0.0550 | 0.0514 |
| | 2000 | 1.0000 | 0.9961 | 0.0514 | 0.0506 |
| | 3000 | 1.0000 | 0.9985 | 0.0553 | 0.0504 |
| | 4000 | 1.0000 | 0.9994 | 0.0489 | 0.0503 |
| | 5000 | 1.0000 | 0.9998 | 0.0467 | 0.0503 |
| | 10000 | 1.0000 | 1.0000 | 0.5060 | 0.0501 |
| Type III | 100 | 1.0000 | 0.3324 | 0.0550 | 0.0521 |
| | 200 | 1.0000 | 0.3606 | 0.0475 | 0.0509 |
| | 500 | 1.0000 | 0.4398 | 0.0503 | 0.0503 |
| | 1000 | 1.0000 | 0.5572 | 0.0533 | 0.0502 |
| | 2000 | 1.0000 | 0.7313 | 0.0503 | 0.0501 |
| | 3000 | 1.0000 | 0.8478 | 0.0472 | 0.0501 |
| | 4000 | 1.0000 | 0.9229 | 0.0494 | 0.0500 |
| | 5000 | 1.0000 | 0.9587 | 0.0467 | 0.0500 |
| | 10000 | 1.0000 | 0.9990 | 0.0478 | 0.0500 |

## Appendix D. QQ Plots

**Figure D.1.** Type II Time-Dependent Covariate — BFGS GMM

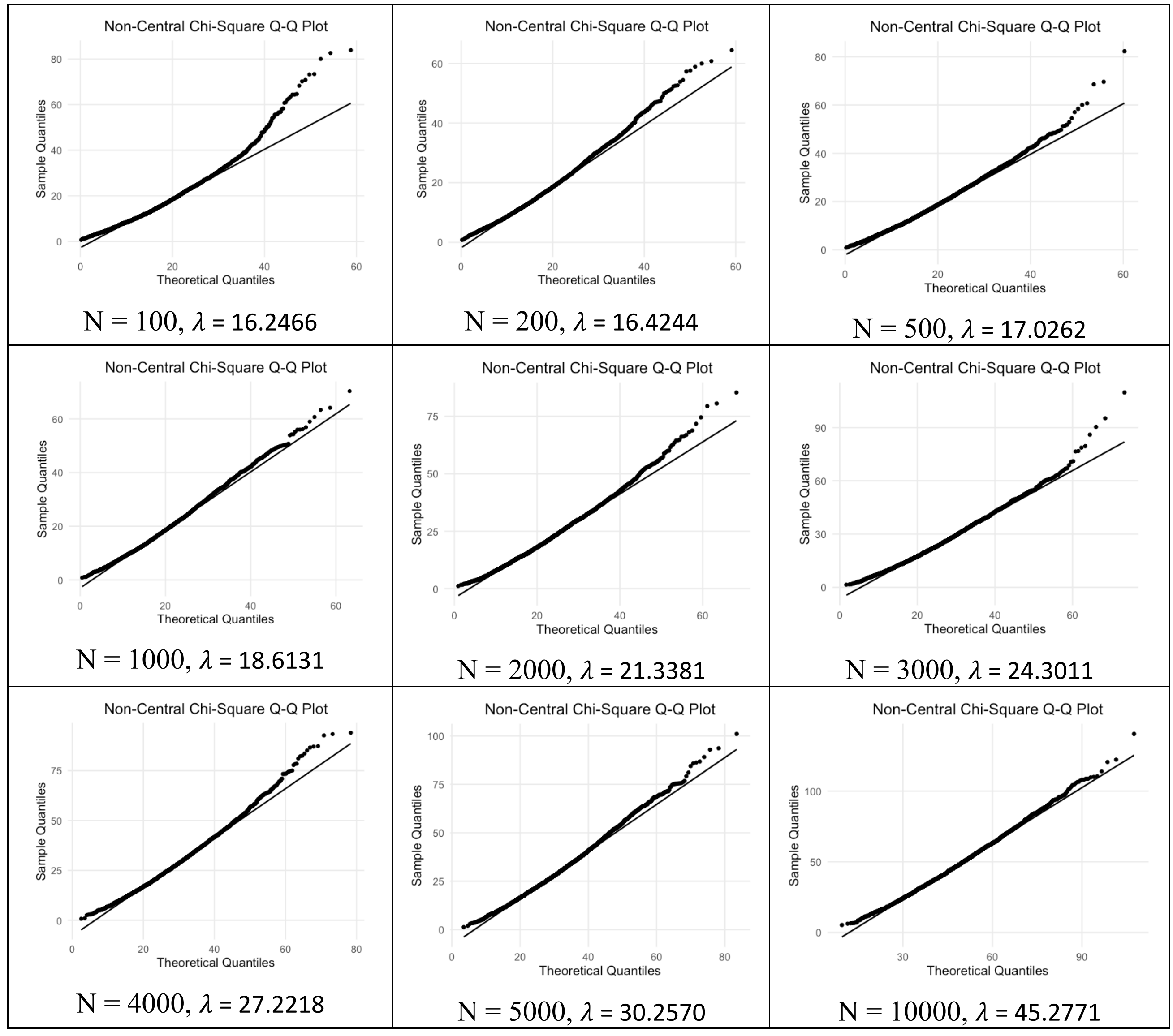

**Figure D.2.** Type II Time-Dependent Covariate — Nelder–Mead GMM

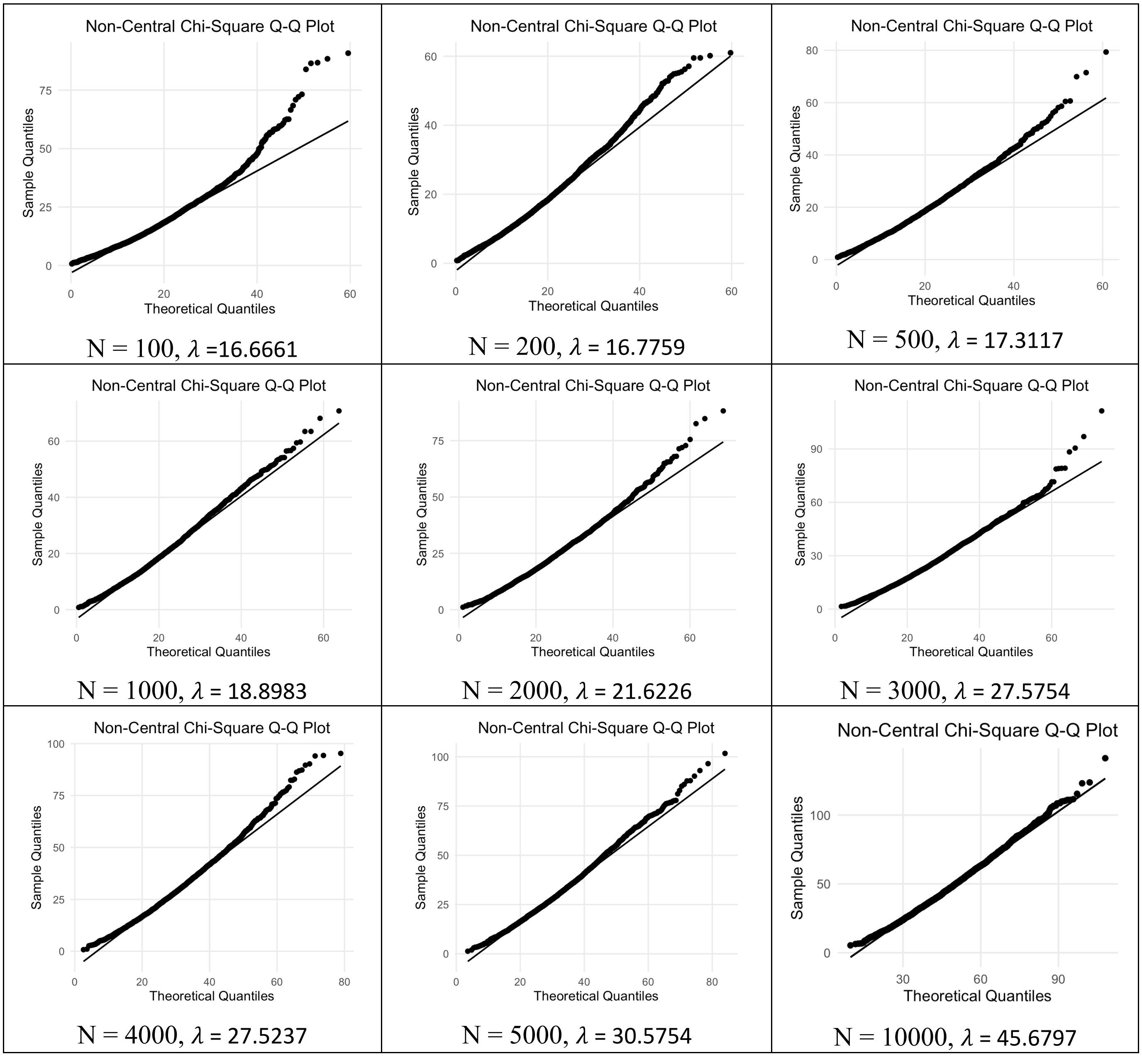

**Figure D.3.** Type III Time-Dependent Covariate — BFGS GMM

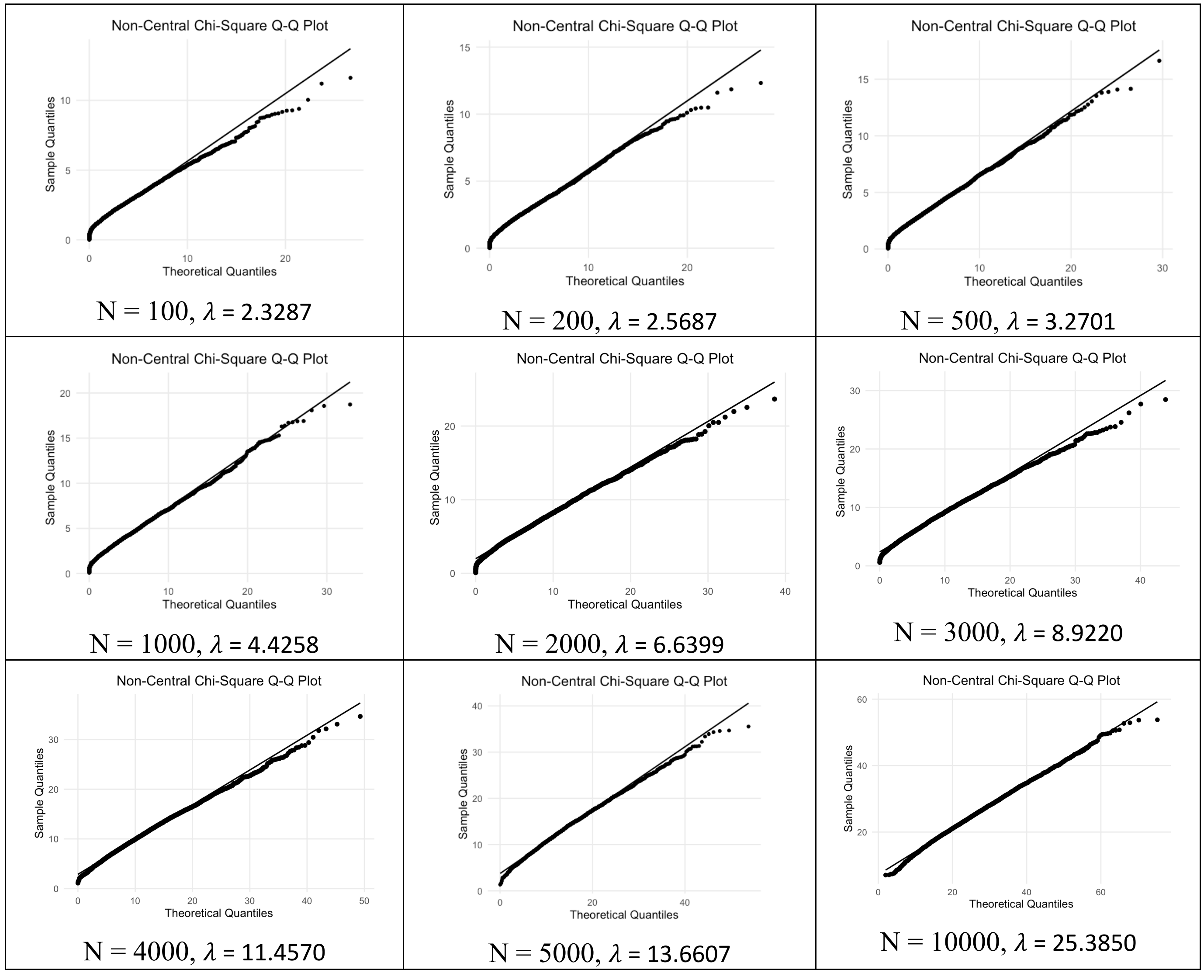

**Figure D.4.** Type III Time-Dependent Covariate — Nelder–Mead GMM

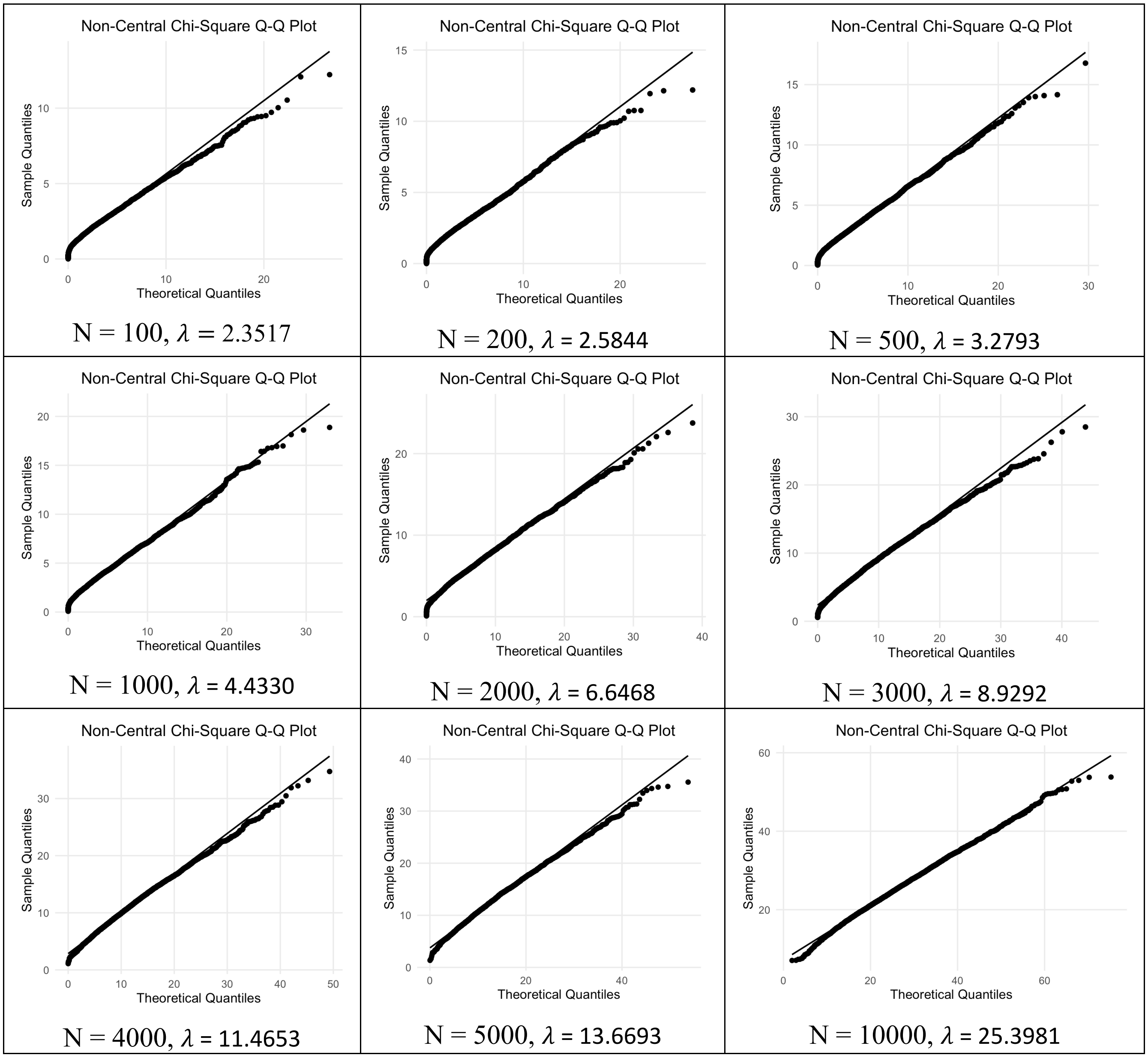